\documentclass[12pt]{article}
\usepackage{amssymb}
\usepackage{graphicx}
\usepackage{amsmath}
\usepackage{doc}

\input{tcilatex}
\begin{document}

\begin{center}
{\Large Exact Solutions for an MHD Generalized Burgers fluid: Stokes' Second
Problem}

\bigskip

\textsl{Masood Khan}\textit{\footnote{%
Corresponding author: Electronic mail: mkhan@qau.edu.pk; mkhan\_21@yahoo.com
(M. Khan)}}$,$ \textsl{Rabia Malik and Asia Anjum}

\textsl{Department of Mathematics, Quaid-i-Azam University, Islamabad }$%
\QTR{sl}{44000}$\textsl{,}

\textsl{Pakistan}

\bigskip

\bigskip
\end{center}

\begin{quote}
\textbf{Abstract: }{\small This paper offers the exact analytical solutions
for the magnetohydrodynamic (MHD) flow of an incompressible generalized
Burgers fluid corresponding to the second problem of Stokes in the presence
of the transverse magnetic field. Modified Darcy's law has been taken into
account. The expression for the velocity field and associated tangential
stress, presented as a sum of the steady-state and transient solutions, are
obtained by means of the integral transforms. Moreover, several figures are
plotted to investigate the effects of various emerging parameters on the
velocity field. The obtained results show that the magnitude of the velocity
and boundary layer thickness significantly reduce in the presence of
magnetic field.}

\end{quote}

\section{Introduction}

Considerable progress has been made in studying flows of non-Newtonian
fluids throughout the last few decades. Due to their viscoelastic nature,
non-Newtonian fluids, such as oils, paints, ketchup, liquid polymers,
asphalt and for forth exhibit some remarkable phenomena. Amplifying interest
of many researchers has shown that these flows are imperative in industry,
manufacturing of food and paper, polymer processing and technology.
Dissimilar to the Newtonian fluid, the flows of non-Newtonian fluids cannot
be explained by a single constitutive model. Therefore, models have been
recognized due to the rheological properties of non-Newtonian fluids.
Amongst them rate type fluid model $[1]$ has received great devotion. These
fluids exhibit the relaxation and retardation phenomena. The simplest
subclasses of rate type fluids are the Maxwell and Oldroyd-B fluids. It is
not an easy task to obtain the analytical solutions for such fluids. In
spite of several challenges, many researchers have established the
analytical solutions regarding these fluids $[2-10].$ However, these fluids
do not predict the rheological properties of some fluids like cheese in food
products and asphalt in geomechanics. In 1935, one-dimensional rate-type
model known as the Burgers model $[11]$ was put in a thermodynamic
framework, which was later extended to the frame-indifferent
three-dimensional form by Krishnan and Rajagopal $[12]$. This model has been
utilized to describe the motion of the earth mental. This model is also the
preferred model to explain the response of asphalt and asphalt concrete $%
[13] $. In addition, Burgers model is sometimes used to model other
geological structures, such as Olivine rocks $[14]$ and the propagation of
seismic waves in the interior of the earth $[15]$. Here, we mention some
studies $[16-20]$ related to the Burgers fluid. Moreover, MHD flow involving
such fluids has promising applications on the development of energy
generation, astrophysics and geophysics fluid dynamics. Recently the theory
of MHD has received great attention $[21,22]$ and references therein. The
effects of transverse magnetic field in the porous space over the unsteady
non-Newtonian fluids are analyzed by several researchers $[23-25].$

Motivated by the above mentioned studies, this work presents an MHD flow of
a generalized Burgers fluid through a porous space due to the oscillation of
an infinite rigid plate. The transverse magnetic field and modified Darcy's
law and their influence on the flow are considered. The solutions are
presented as a sum of steady-state and transient solutions. The effects of
some physical parameters are discussed through graphical illustrations.

\section{Governing Equations}

The Cauchy stress tensor $\mathbf{T}$ in a generalized Burgers\textbf{\ }%
fluid is given by $[16-20]$

\begin{equation}
\mathbf{T}=-p\mathbf{I}+\mathbf{S,}\text{ \ \ \ \ \ \ \ \ \ }\mathbf{S}%
+\lambda _{1}\frac{\delta \mathbf{S}}{\delta t}+\lambda _{2}\frac{\delta ^{2}%
\mathbf{S}}{\delta t^{2}}=\mu \left[ \mathbf{A}_{1}+\lambda _{3}\frac{\delta 
\mathbf{A}_{1}}{\delta t}+\lambda _{4}\frac{\delta ^{2}\mathbf{A}_{1}}{%
\delta t^{2}}\right] ,  \tag{1}
\end{equation}%
where $\mathbf{S}$ is the extra-stress tensor, $p$ the pressure, $\mathbf{I}$
the identity tensor, $\mu $ the dynamic viscosity, $\mathbf{A}_{1}=\mathbf{L}%
+\mathbf{L}^{\text{\textsf{T}}}$ the first Rivlin-Ericksen tensor with $%
\mathbf{L}$ as the velocity gradient, $\lambda _{1}$ and $\lambda _{3}$\ $%
(\leq \lambda _{1})$ are relaxation and retardation times, respectively, $%
\lambda _{2}$ \ and $\lambda _{4}$ are material parameters of generalized
Burgers fluid and $\delta /\delta t$ denotes the upper convected time
derivative defined by

\begin{equation}
\frac{\delta \mathbf{S}}{\delta t}=\frac{d\mathbf{S}}{dt}-\mathbf{LS}-%
\mathbf{SL}^{\text{\textsf{T}}},\text{ \ \ \ }\frac{\delta ^{2}\mathbf{S}}{%
\delta t^{2}}=\frac{\delta }{\delta t}\left( \frac{\delta \mathbf{S}}{\delta
t}\right) ,  \tag{2}
\end{equation}%
in which $d/dt$ is the material time derivative.

The basic equations governing the unsteady flow of an incompressible fluid
are%
\begin{equation}
\func{div}\mathbf{V=}0\mathbf{,}  \tag{3}
\end{equation}%
\begin{equation}
\rho \frac{d\mathbf{V}}{dt}=\func{div}\mathbf{T\mathbf{-}}\sigma B_{0}^{2}%
\mathbf{\mathbf{V}+R,}  \tag{4}
\end{equation}%
where $\mathbf{V}$\textbf{\ }is the velocity, $\rho $ the density of the
fluid, $\sigma $\ the electrical conductivity of the fluid, $B_{0}$ the
magnitude of applied magnetic field and $\mathbf{R}$ denotes the Darcy's
resistance.

For the problem under consideration we shall assume the velocity and stress
fields of the form

\begin{equation}
\mathbf{V}=\mathbf{V}(y,t)=u(y,t)\mathbf{i},\text{ \ \ \ \ }\mathbf{S}=%
\mathbf{S}(y,t),  \tag{5}
\end{equation}%
where $\mathbf{i}$ is the unit vector along the $x-$coordinate direction.
The velocity field $(5)$ automatically satisfies the continuity equation $%
(3).$

Substitution of Eq. $(5)$ in Eq. $(1)$ and having in mind the initial
conditions 
\begin{equation}
\mathbf{S}(y,0)=\frac{\partial \mathbf{S}(y,0)}{\partial t}=\mathbf{0,} 
\tag{6}
\end{equation}%
yields $S_{yy}=S_{yz}=S_{zz}=S_{xz}=0$ and%
\begin{equation}
\left( 1+\lambda _{1}\frac{\partial }{\partial t}+\lambda _{2}\frac{\partial
^{2}}{\partial t^{2}}\right) S_{xy}=\mu \left( 1+\lambda _{3}\frac{\partial 
}{\partial t}+\lambda _{4}\frac{\partial ^{2}}{\partial t^{2}}\right) \frac{%
\partial u}{\partial y},  \tag{7}
\end{equation}%
where $S_{xy}$ is the tangential stress.

In view of the reference $[22]$, we have the following relation of $\mathbf{R%
}$ for a generalized Burgers fluid%
\begin{equation}
\left( 1+\lambda _{1}\frac{\partial }{\partial t}+\lambda _{2}\frac{\partial
^{2}}{\partial t^{2}}\right) \mathbf{R=-}\frac{\mu \varphi }{k}\left(
1+\lambda _{3}\frac{\partial }{\partial t}+\lambda _{4}\frac{\partial ^{2}}{%
\partial t^{2}}\right) \mathbf{V}(y,t)\mathbf{,}  \tag{8}
\end{equation}%
where $\varphi $ is the porosity and $k$ the permeability of the porous
medium.

By substituting Eq. $(5)$ in the balance of linear momentum $(4)$, having in
mind Eqs. $(7)$ and $(8)$ and assuming that there is no pressure gradient in
the flow direction, one finds the following governing equation

\begin{eqnarray}
\left( 1+\lambda _{1}\frac{\partial }{\partial t}+\lambda _{2}\frac{\partial
^{2}}{\partial t^{2}}\right) \frac{\partial u(y,t)}{\partial t} &=&\nu
\left( 1+\lambda _{3}\frac{\partial }{\partial t}+\lambda _{4}\frac{\partial
^{2}}{\partial t^{2}}\right) \frac{\partial ^{2}u(y,t)}{\partial y^{2}} 
\notag \\
&&-\frac{\sigma B_{0}^{2}}{\rho }\left( 1+\lambda _{1}\frac{\partial }{%
\partial t}+\lambda _{2}\frac{\partial ^{2}}{\partial t^{2}}\right) u(y,t)%
\mathbf{-}\frac{\nu \varphi }{k}\left( 1+\lambda _{3}\frac{\partial }{%
\partial t}+\lambda _{4}\frac{\partial ^{2}}{\partial t^{2}}\right) u(y,t), 
\notag \\
&&  \TCItag{9}
\end{eqnarray}%
where $\nu =\mu /\rho $\ is the kinematic viscosity of the fluid.

\section{Statement of the Problem and its Solution}

Let we consider an incompressible and electrically conducting generalized
Burgers fluid. The fluid occupies the porous space above the flat plate
perpendicular to the $y-$axis and permeated by an applied magnetic field $%
B_{0}$ normal to the flow. For $t>0$ the plate starts to oscillate in its
own plane with velocity $UH(t)\cos (wt)$ or $UH(t)\sin (wt)$ with $H(\cdot )$
as the Heaviside unit step function and $U$ the amplitude of the velocity of
the plate. Due to the shear, the fluid above the plate is gradually moved.
The governing equation of the problem is (9) and the associated initial and
boundary conditions are%
\begin{equation}
u(y,0)=\frac{\partial u(y,0)}{\partial t}=\frac{\partial ^{2}u(y,0)}{%
\partial t^{2}}=0;\text{ \ }y>0,\text{ \ }  \tag{10}
\end{equation}%
\begin{equation}
u(0,t)=UH(t)\cos \left( wt\right) \text{ or }u(0,t)=UH(t)\sin \left(
wt\right) ;\text{ }t>0\text{ \ },  \tag{11}
\end{equation}

and%
\begin{equation}
u(y,t),\frac{\partial u(y,t)}{\partial y}\rightarrow 0\ \text{as }%
y\rightarrow \infty \text{ and }t>0.  \tag{12}
\end{equation}

In the non-dimensional form we can write the above problem as

\begin{eqnarray}
\left( 1+\frac{\partial }{\partial t}+\beta \frac{\partial ^{2}}{\partial
t^{2}}\right) \frac{\partial u(y,t)}{\partial t} &=&\nu \left( 1+\alpha 
\frac{\partial }{\partial t}+\gamma \frac{\partial ^{2}}{\partial t^{2}}%
\right) \frac{\partial ^{2}u(y,t)}{\partial y^{2}}-M^{2}\left( 1+\frac{%
\partial }{\partial t}+\beta \frac{\partial ^{2}}{\partial t^{2}}\right)
u(y,t)  \notag \\
&&\mathbf{-}\frac{1}{K}\left( 1+\alpha \frac{\partial }{\partial t}+\gamma 
\frac{\partial ^{2}}{\partial t^{2}}\right) u(y,t),  \TCItag{13}
\end{eqnarray}%
\begin{equation}
u(y,0)=\frac{\partial u(y,0)}{\partial t}=\frac{\partial ^{2}u(y,0)}{%
\partial t^{2}}=0;\text{ \ }y>0,\text{ \ }  \tag{14}
\end{equation}%
\begin{equation}
u(0,t)=H(t)\cos \left( wt\right) \text{ or }u(0,t)=H(t)\sin \left( wt\right)
;\text{ }t>0\text{ \ },  \tag{15}
\end{equation}%
and%
\begin{equation}
u(y,t),\frac{\partial u(y,t)}{\partial y}\rightarrow 0\ \text{as }%
y\rightarrow \infty \text{ and }t>0,  \tag{16}
\end{equation}%
where

\begin{equation}
u^{\ast }=\frac{u}{U},t^{\ast }=\frac{t}{\lambda _{1}},y^{\ast }=\frac{y}{%
\sqrt{\nu \lambda _{1}}},\alpha =\frac{\lambda _{3}}{\lambda _{1}},\beta =%
\frac{\lambda _{2}}{\lambda _{1}^{2}},\gamma =\frac{\lambda _{4}}{\lambda
_{1}^{2}},M^{2}=\frac{\sigma B_{0}^{2}\lambda _{1}}{\rho },\frac{1}{K}=\frac{%
\lambda _{1}\nu \varphi }{k},  \tag{17}
\end{equation}%
with asterisks have been omitted for simplicity.

\subsection{Calculation of the Velocity Field}

In order to obtain the exact solution describing the flow for small and
large times, we shall use the Fourier sine and Laplace transforms $[26].$
Thus, multiplying both sides of Eq. $(13)$ by $\sin (y\xi )$ integrating the
result with respect to $y$ from $0$ to infinity, and taking into account the
initial and boundary conditions $(14)-(16)$, we find that%
\begin{eqnarray}
&&\beta \frac{\partial ^{3}u_{s}(\xi ,t)}{\partial t^{3}}+\left( 1+\xi
^{2}\gamma +\beta M^{2}+\frac{\gamma }{K}\right) \frac{\partial
^{2}u_{s}(\xi ,t)}{\partial t^{2}}  \notag \\
&&+\left( 1+\alpha \xi ^{2}+M^{2}+\frac{\alpha }{K}\right) \frac{\partial
u_{s}(\xi ,t)}{\partial t}+\left( \xi ^{2}+M^{2}+\frac{1}{K}\right)
u_{s}(\xi ,t)  \notag \\
&=&\left( 1+\alpha \frac{\partial }{\partial t}+\gamma \frac{\partial ^{2}}{%
\partial t^{2}}\right) \xi H(t)\cos (wt);\text{ \ }\xi ,t>0,  \TCItag{18}
\end{eqnarray}%
respectively,

\begin{eqnarray}
&&\beta \frac{\partial ^{3}u_{s}(\xi ,t)}{\partial t^{3}}+\left( 1+\xi
^{2}\gamma +\beta M^{2}+\frac{\gamma }{K}\right) \frac{\partial
^{2}u_{s}(\xi ,t)}{\partial t^{2}}  \notag \\
&&+\left( 1+\alpha \xi ^{2}+M^{2}+\frac{\alpha }{K}\right) \frac{\partial
u_{s}(\xi ,t)}{\partial t}+\left( \xi ^{2}+M^{2}+\frac{1}{K}\right)
u_{s}(\xi ,t)  \notag \\
&=&\left( 1+\alpha \frac{\partial }{\partial t}+\gamma \frac{\partial ^{2}}{%
\partial t^{2}}\right) \xi H(t)\sin (wt);\text{ \ }\xi ,t>0,  \TCItag{19}
\end{eqnarray}
where the Fourier sine transform has to satisfy the initial conditions%
\begin{equation}
u_{s}(\xi ,0)=\frac{\partial u_{s}(\xi ,0)}{\partial t}=\frac{\partial
^{2}u_{s}(\xi ,0)}{\partial t^{2}}=0\text{; \ \ }\xi >0\text{.}  \tag{20}
\end{equation}

Now taking the Laplace transform of Eqs. $(18)$ and $(19)$ subject to the
initial conditions $(20),$ we obtain%
\begin{equation}
\bar{U}_{s}\left( \xi ,q\right) =\frac{\xi }{\beta }\frac{\left( 1+\alpha
q+\gamma q^{2}\right) q^{2}}{q(q^{2}+w^{2})\left[ q^{3}+\left( \frac{1+\xi
^{2}\gamma +a_{m}}{\beta }\right) q^{2}+\left( \frac{1+\alpha \xi ^{2}+c_{m}%
}{\beta }\right) q+\left( \frac{\xi ^{2}+b_{m}}{\beta }\right) \right] }, 
\tag{21}
\end{equation}%
respectively,

\begin{equation}
\bar{U}_{s}\left( \xi ,q\right) =\frac{\xi }{\beta }\frac{w\left( 1+\alpha
q+\gamma q^{2}\right) q}{q(q^{2}+w^{2})\left[ q^{3}+\left( \frac{1+\xi
^{2}\gamma +a_{m}}{\beta }\right) q^{2}+\left( \frac{1+\alpha \xi ^{2}+c_{m}%
}{\beta }\right) q+\left( \frac{\xi ^{2}+b_{m}}{\beta }\right) \right] }, 
\tag{22}
\end{equation}%
where $\bar{U}_{s}\left( \xi ,q\right) $ is the Laplace transform of $%
u_{s}(\xi ,t),$ $q$ the transform parameter and

\begin{equation*}
a_{m}=\beta M^{2}+\frac{\gamma }{K},\text{ \ \ \ \ }b_{m}=M^{2}+\frac{1}{K},%
\text{ \ \ \ }c_{m}=M^{2}+\frac{\alpha }{K}.
\end{equation*}

Rewriting Eqs. $(21)$ and $(22)$ in simpler form as:

\begin{equation}
\bar{U}_{s}\left( \xi ,q\right) =\frac{\xi }{\beta }\frac{G(\xi ,q)}{q}, 
\tag{23}
\end{equation}%
where

\begin{equation}
G(\xi ,q)=\frac{\left( 1+\alpha q+\gamma q^{2}\right) q^{2}}{%
(q^{2}+w^{2})(q-q_{1})(q-q_{2})(q-q_{3})},  \tag{24}
\end{equation}%
respectively%
\begin{equation}
G(\xi ,q)=\frac{w\left( 1+\alpha q+\gamma q^{2}\right) q}{%
(q^{2}+w^{2})(q-q_{1})(q-q_{2})(q-q_{3})}.  \tag{25}
\end{equation}%
Equations $(24)$ and $(25)$ can also be written as

\begin{eqnarray}
G(\xi ,q) &=&\frac{q_{1}^{2}(1+\alpha q_{1}+\gamma q_{1}^{2})}{%
(q_{1}-q_{2})(q_{1}-q_{3})(q_{1}^{2}+w^{2})}\frac{1}{(q-q_{1})}-\frac{%
q_{2}^{2}(1+\alpha q_{2}+\gamma q_{2}^{2})}{%
(q_{1}-q_{2})(q_{2}-q_{3})(q_{2}^{2}+w^{2})}\frac{1}{(q-q_{2})}  \notag \\
&&+\frac{q_{3}^{2}(1+\alpha q_{3}+\gamma q_{3}^{2})}{%
(q_{1}-q_{3})(q_{2}-q_{3})(q_{3}^{2}+w^{2})}\frac{1}{(q-q_{3})}+\frac{\psi
_{1}}{\eta }\frac{w}{(q^{2}+w^{2})}+\frac{\psi _{2}}{\eta }\frac{q}{%
(q^{2}+w^{2})},  \TCItag{26}
\end{eqnarray}%
respectively,%
\begin{eqnarray}
G(\xi ,q) &=&\frac{wq_{1}(1+\alpha q_{1}+\gamma q_{1}^{2})}{%
(q_{1}-q_{2})(q_{1}-q_{3})(q_{1}^{2}+w^{2})}\frac{1}{(q-q_{1})}-\frac{%
wq_{2}(1+\alpha q_{2}+\gamma q_{2}^{2})}{%
(q_{1}-q_{2})(q_{2}-q_{3})(q_{2}^{2}+w^{2})}\frac{1}{(q-q_{2})}  \notag \\
&&+\frac{wq_{3}(1+\alpha q_{3}+\gamma q_{3}^{2})}{%
(q_{1}-q_{3})(q_{2}-q_{3})(q_{3}^{2}+w^{2})}\frac{1}{(q-q_{3})}+\frac{\psi
_{2}}{\eta }\frac{w}{(q^{2}+w^{2})}-\frac{\psi _{1}}{\eta }\frac{q}{%
(q^{2}+w^{2})},  \TCItag{27}
\end{eqnarray}

where

\begin{eqnarray*}
q_{j} &=&s_{j}-\frac{1+\xi ^{2}\gamma +a_{m}}{3\beta ^{2}};\text{ \ \ \ }%
j=1,2,3, \\
s_{1} &=&\sqrt[3]{-\frac{p_{1}}{2}+\sqrt{\frac{p_{1}^{2}}{4}+\frac{p_{2}^{3}%
}{27}}}+\sqrt[3]{-\frac{p_{1}}{2}-\sqrt{\frac{p_{1}^{2}}{4}+\frac{p_{2}^{3}}{%
27}}}, \\
s_{2} &=&h\sqrt[3]{-\frac{p_{1}}{2}+\sqrt{\frac{p_{1}^{2}}{4}+\frac{p_{2}^{3}%
}{27}}}+h^{2}\sqrt[3]{-\frac{p_{1}}{2}-\sqrt{\frac{p_{1}^{2}}{4}+\frac{%
p_{2}^{3}}{27}},} \\
s_{3} &=&h^{2}\sqrt[3]{-\frac{p_{1}}{2}+\sqrt{\frac{p_{1}^{2}}{4}+\frac{%
p_{2}^{3}}{27}}}+h\sqrt[3]{-\frac{p_{1}}{2}-\sqrt{\frac{p_{1}^{2}}{4}+\frac{%
p_{2}^{3}}{27}}}, \\
\text{\ \ }p_{1} &=&\frac{\xi ^{2}+b_{m}}{\beta }-\frac{\left( 1+\xi
^{2}\gamma +a_{m}\right) \left( 1+\alpha \xi ^{2}+c_{m}\right) }{3\beta ^{2}}%
+2\frac{\left( 1+\xi ^{2}\gamma +a_{m}\right) ^{3}}{27\beta ^{3}}, \\
p_{2} &=&\frac{1+\alpha \xi ^{2}+c_{m}}{\beta }-\frac{\left( 1+\xi
^{2}\gamma +a_{m}\right) ^{2}}{3\beta ^{2}},\text{ \ }h=\frac{-1+i\sqrt{3}}{2%
},\text{ } \\
\psi _{1} &=&q_{1}q_{2}+q_{1}q_{3}+q_{2}q_{3}-w^{2}+q_{1}q_{2}q_{3}\alpha
-q_{1}w^{2}\alpha -q_{2}w^{2}\alpha -q_{3}w^{2}\alpha -q_{1}q_{2}w^{2}\gamma
\\
&&-q_{1}q_{3}w^{2}\gamma -q_{2}q_{3}w^{2}\gamma +w^{4}\gamma , \\
&& \\
\psi _{2}
&=&-q_{1}q_{2}q_{3}+q_{1}w^{2}+q_{2}w^{2}+q_{3}w^{2}+q_{1}q_{2}w^{2}\alpha
+q_{1}q_{3}w^{2}\alpha +q_{2}q_{3}w^{2}\alpha -w^{4}\alpha \\
&&+q_{1}q_{2}q_{3}w^{2}\gamma -q_{1}w^{4}\gamma -q_{2}w^{4}\gamma
-q_{3}w^{4}\gamma , \\
&& \\
\psi _{3} &=&(q_{1}^{2}+w^{2})(q_{2}^{2}+w^{2})(q_{3}^{2}+w^{2}).
\end{eqnarray*}%
Applying the inverse Laplace transform to Eq. $(23)$, one obtains

\begin{equation}
u_{s}(\xi ,t)=\frac{\xi }{\beta }\left[ \frac{\chi _{1}e^{q_{1}t}}{%
(q_{1}-q_{2})(q_{1}-q_{3})}-\frac{\chi _{2}e^{q_{2}t}}{%
(q_{1}-q_{2})(q_{2}-q_{3})}+\frac{\chi _{3}e^{q_{3}t}}{%
(q_{1}-q_{3})(q_{2}-q_{3})}+\frac{\psi _{1}}{\eta }\sin \left( wt\right) +%
\frac{\psi _{2}}{\eta }\cos \left( wt\right) \right] ,  \tag{28}
\end{equation}%
respectively,%
\begin{equation}
u_{s}(\xi ,t)=\frac{\xi }{\beta }\left[ \frac{\chi _{1}^{\ast }e^{q_{1}t}}{%
(q_{1}-q_{2})(q_{1}-q_{3})}-\frac{\chi _{2}^{\ast }e^{q_{2}t}}{%
(q_{1}-q_{2})(q_{2}-q_{3})}+\frac{\chi _{3}^{\ast }e^{q_{3}t}}{%
(q_{1}-q_{3})(q_{2}-q_{3})}+\frac{\psi _{2}}{\eta }\sin \left( wt\right) -%
\frac{\psi _{1}}{\eta }\cos \left( wt\right) \right] ,  \tag{29}
\end{equation}%
where%
\begin{equation*}
\chi _{i}=\frac{q_{i}(1+\alpha q_{i}+\gamma q_{i}^{2})}{(q_{i}^{2}+w^{2})},%
\text{ \ \ \ \ \ \ }\chi _{i}^{\ast }=\frac{w(1+\alpha q_{i}+\gamma
q_{i}^{2})}{(q_{i}^{2}+w^{2})},i=1,2,3.
\end{equation*}

Inverting Eqs. $(28)$ and $(29)$\ by means of the Fourier's sine formulae $%
[26],$ we can write the starting solutions\ as%
\begin{eqnarray}
u(y,t) &=&H(t)\frac{1}{\beta }\int_{0}^{\infty }\left( \frac{\chi
_{1}e^{q_{1}t}}{(q_{1}-q_{2})(q_{1}-q_{3})}-\frac{\chi _{2}e^{q_{2}t}}{%
(q_{1}-q_{2})(q_{2}-q_{3})}+\frac{\chi _{3}e^{q_{3}t}}{%
(q_{1}-q_{3})(q_{2}-q_{3})}\right) \xi \sin (\xi y)d\xi  \notag \\
&&+\frac{\psi _{1}}{\beta X}H(t)\sin \left( wt\right) \int_{0}^{\infty }%
\frac{\xi \sin (\xi y)}{\left( \xi ^{2}+Y^{2}\right) ^{2}+Z^{2}}d\xi
+H(t)\cos \left( wt\right) \int_{0}^{\infty }\frac{\left( \xi
^{2}+Y^{2}\right) \sin (\xi y)}{\left( \xi ^{2}+Y^{2}\right) ^{2}+Z^{2}}d\xi
,  \notag \\
&&  \TCItag{30}
\end{eqnarray}%
respectively,%
\begin{eqnarray}
u(y,t) &=&\frac{H(t)}{\beta }\int_{0}^{\infty }\left( \frac{\chi _{1}^{\ast
}e^{q_{1}t}}{(q_{1}-q_{2})(q_{1}-q_{3})}-\frac{\chi _{2}^{\ast }e^{q_{2}t}}{%
(q_{1}-q_{2})(q_{2}-q_{3})}+\frac{\chi _{3}^{\ast }e^{q_{3}t}}{%
(q_{1}-q_{3})(q_{2}-q_{3})}\right) \xi \sin (\xi y)d\xi  \notag \\
&&+H(t)\sin \left( wt\right) \int_{0}^{\infty }\frac{\left( \xi
^{2}+Y^{2}\right) \sin (\xi y)}{\left( \xi ^{2}+Y^{2}\right) ^{2}+Z^{2}}d\xi
-\frac{\psi _{1}}{\beta X}H(t)\cos \left( wt\right) \int_{0}^{\infty }\frac{%
\xi \sin (\xi y)}{\left( \xi ^{2}+Y^{2}\right) ^{2}+Z^{2}}d\xi ,  \notag \\
&&  \TCItag{31}
\end{eqnarray}%
with

\begin{eqnarray*}
X &=&\frac{1+(\alpha ^{2}-2\gamma )w^{2}+\gamma ^{2}w^{4}}{\beta ^{2}}, \\
Y^{2} &=&\frac{\left[ \gamma \left( 1+a_{m}\right) -\alpha \beta \right]
w^{4}+\left[ \alpha \left( 1+c_{m}\right) -\left( 1+a_{m}+\gamma
b_{m}\right) \right] w^{2}+bm}{1+(\alpha ^{2}-2\gamma )w^{2}+\gamma ^{2}w^{4}%
}, \\
Z^{2} &=&\frac{w^{2}\left[ 1+c_{m}-b_{m}\alpha -c_{m}w^{2}\gamma
+w^{2}\left( \alpha +a_{m}\alpha -\beta -\gamma +w^{2}\beta \gamma \right) %
\right] ^{2}}{\left[ 1+w^{2}\left\{ \alpha ^{2}+\gamma \left( -2+w^{2}\gamma
\right) \right\} \right] ^{2}}.
\end{eqnarray*}

As we know the following relations%
\begin{equation}
\int_{0}^{\infty }\frac{\xi \sin (\xi y)}{\left( \xi ^{2}+Y^{2}\right)
^{2}+Z^{2}}d\xi =\frac{\pi }{2Z}e^{-Ay}\sin \left( By\right) ,\text{ \ \ }%
\int_{0}^{\infty }\frac{\left( \xi ^{2}+Y^{2}\right) \sin (\xi y)}{\left(
\xi ^{2}+Y^{2}\right) ^{2}+Z^{2}}d\xi =\frac{\pi }{2}e^{-Ay}\cos \left(
By\right) ,  \tag{32}
\end{equation}%
where%
\begin{equation*}
2A^{2}=\sqrt{Y^{4}+Z^{2}}+Y^{2},\text{ \ \ \ }2B^{2}=\sqrt{Y^{4}+Z^{2}}%
-Y^{2}.
\end{equation*}

Using the above relations into Eqs. $(30)$ and $(31)$ we get the following
simplified expressions for the starting solutions

\begin{eqnarray}
u(y,t) &=&\frac{H(t)}{\beta }\int_{0}^{\infty }\left( \frac{\chi
_{1}e^{q_{1}t}}{(q_{1}-q_{2})(q_{1}-q_{3})}-\frac{\chi _{2}e^{q_{2}t}}{%
(q_{1}-q_{2})(q_{2}-q_{3})}+\frac{\chi _{3}e^{q_{3}t}}{%
(q_{1}-q_{3})(q_{2}-q_{3})}\right) \xi \sin (\xi y)d\xi  \notag \\
&&+\frac{\pi }{2}H(t)e^{-Ay}\cos (wt-By),  \TCItag{33}
\end{eqnarray}%
respectively,%
\begin{eqnarray}
u(y,t) &=&\frac{H(t)}{\beta }\int_{0}^{\infty }\left( \frac{\chi _{1}^{\ast
}e^{q_{1}t}}{(q_{1}-q_{2})(q_{1}-q_{3})}-\frac{\chi _{2}^{\ast }e^{q_{2}t}}{%
(q_{1}-q_{2})(q_{2}-q_{3})}+\frac{\chi _{3}^{\ast }e^{q_{3}t}}{%
(q_{1}-q_{3})(q_{2}-q_{3})}\right) \xi \sin (\xi y)d\xi  \notag \\
&&+\frac{\pi }{2}H(t)e^{-Ay}\sin (wt-By).  \TCItag{34}
\end{eqnarray}

The starting solutions $(33)$ and $(34)$ are presented as a sum of the
steady-state and transient solutions. The steady-state solutions are

\begin{equation}
u(y,t)=\frac{\pi }{2}H(t)e^{-Ay}\cos (wt-By),  \tag{35}
\end{equation}%
respectively,%
\begin{equation}
u(y,t)=\frac{\pi }{2}H(t)e^{-Ay}\sin (wt-By).  \tag{36}
\end{equation}

\subsection{Calculation of the Shear Stress}

To determine the shear stress, we use relation $\left( 7\right) $ in the
non-dimensional form%
\begin{equation}
\left( 1+\frac{\partial }{\partial t}+\beta \frac{\partial ^{2}}{\partial
t^{2}}\right) \tau =\left( 1+\alpha \frac{\partial }{\partial t}+\gamma 
\frac{\partial ^{2}}{\partial t^{2}}\right) \frac{\partial u}{\partial y}. 
\tag{37}
\end{equation}

Applying the Laplace transform to Eq. $(37)$ one obtains%
\begin{equation}
\bar{\tau}(y,q)=\frac{(1+\alpha q+\gamma q^{2})}{\beta (q-q_{4})(q-q_{5})}%
\frac{\partial \bar{U}(y,q)}{\partial y},  \tag{38}
\end{equation}%
where%
\begin{equation*}
q_{4}=\frac{-1+\sqrt{1-4\beta }}{2\beta },\text{ \ \ }q_{5}=\frac{-1-\sqrt{%
1-4\beta }}{2\beta },
\end{equation*}%
and $\bar{\tau}(y,q)$ is the Laplace transform of $\tau (y,t)$ and the image
function $\bar{U}\left( y,q\right) =\tciLaplace \lbrack u(y,t)]$ has been
obtained through Eqs. $(21)$ and $(22)$ by applying the inverse Fourier sine
transform and gets%
\begin{equation}
\bar{U}\left( y,q\right) =\frac{2}{\pi \beta }\int_{0}^{\infty }\frac{\left(
1+\alpha q+\gamma q^{2}\right) q^{2}}{%
q(q^{2}+w^{2})(q-q_{1})(q-q_{2})(q-q_{3})}\xi \sin (\xi y)d\xi ,  \tag{39}
\end{equation}%
respectively%
\begin{equation}
\bar{U}\left( y,q\right) =\frac{2}{\pi \beta }\int_{0}^{\infty }\frac{\left(
1+\alpha q+\gamma q^{2}\right) qw}{q(q^{2}+w^{2})(q-q_{1})(q-q_{2})(q-q_{3})}%
\xi \sin (\xi y)d\xi .  \tag{40}
\end{equation}

Substituting Eqs. $(39)$ and $(40)$ in Eq. $(38),$ we reach at the following
expressions%
\begin{equation}
\bar{\tau}(y,q)=\frac{2}{\pi \beta ^{2}}\int_{0}^{\infty }\frac{(1+\alpha
q+\gamma q^{2})q^{2}}{%
q(q^{2}+w^{2})(q-q_{1})(q-q_{2})(q-q_{3})(q-q_{4})(q-q_{5})}\xi ^{2}\cos
(\xi y)d\xi ,  \tag{41}
\end{equation}%
respectively%
\begin{equation}
\bar{\tau}(y,q)=\frac{2}{\pi \beta ^{2}}\int_{0}^{\infty }\frac{(1+\alpha
q+\gamma q^{2})qw}{%
q(q^{2}+w^{2})(q-q_{1})(q-q_{2})(q-q_{3})(q-q_{4})(q-q_{5})}\xi ^{2}\cos
(\xi y)d\xi ,  \tag{42}
\end{equation}%
or in simpler form we can write as%
\begin{equation}
\bar{\tau}(y,q)=\frac{2}{\pi \beta ^{2}}\int_{0}^{\infty }\frac{F(\xi ,q)}{q}%
\xi ^{2}\cos (\xi y)d\xi ,  \tag{43}
\end{equation}%
where%
\begin{equation}
F(\xi ,q)=\frac{(1+\alpha q+\gamma q^{2})q^{2}}{%
(q^{2}+w^{2})(q-q_{1})(q-q_{2})(q-q_{3})(q-q_{4})(q-q_{5})},  \tag{44}
\end{equation}%
respectively%
\begin{equation}
F(\xi ,q)=\frac{(1+\alpha q+\gamma q^{2})qw}{%
(q^{2}+w^{2})(q-q_{1})(q-q_{2})(q-q_{3})(q-q_{4})(q-q_{5})}.  \tag{45}
\end{equation}

Inverting Eq. $(43)$ by means of the Laplace transform, following the same
way as for the velocity field, we find the following expressions for the
tangential stress

\begin{eqnarray}
\tau (y,t) &=&\frac{2H(t)}{\pi \beta ^{2}}\int_{0}^{\infty }\left( \frac{%
\zeta _{1}e^{q_{1}t}}{(q_{1}-q_{2})(q_{1}-q_{3})}+\frac{\zeta _{2}e^{q_{2}t}%
}{(q_{2}-q_{1})(q_{2}-q_{3})}+\frac{\zeta _{3}e^{q_{3}t}}{%
(q_{1}-q_{3})(q_{2}-q_{3})}\right) \xi ^{2}\cos (\xi y)d\xi  \notag \\
&&-\frac{H(t)}{\beta ^{2}}e^{-By}\left[ 
\begin{array}{c}
\left[ A\sin \left( Ay\right) -B\cos \left( Ay\right) \right] \left[ \psi
_{4}\cos \left( wt\right) +\psi _{5}\sin \left( wt\right) \right] \\ 
\\ 
-\left[ A\cos \left( Ay\right) -B\sin \left( Ay\right) \right] \left[ \psi
_{6}\cos \left( wt\right) +\psi _{7}\sin \left( wt\right) \right]%
\end{array}%
\right] ,  \TCItag{46}
\end{eqnarray}%
respectively%
\begin{eqnarray}
\tau (y,t) &=&\frac{2H(t)}{\pi \beta ^{2}}\int_{0}^{\infty }\left( \frac{%
\zeta _{1}^{\ast }e^{q_{1}t}}{(q_{1}-q_{2})(q_{1}-q_{3})}+\frac{\zeta
_{2}^{\ast }e^{q_{2}t}}{(q_{2}-q_{1})(q_{2}-q_{3})}+\frac{\zeta _{3}^{\ast
}e^{q_{3}t}}{(q_{1}-q_{3})(q_{2}-q_{3})}\right) \xi ^{2}\cos (\xi y)d\xi 
\notag \\
&&+\frac{H(t)}{\beta ^{2}}e^{-By}\left[ 
\begin{array}{c}
\left[ A\sin \left( Ay\right) -B\cos \left( Ay\right) \right] \left[ \psi
_{4}\cos \left( wt\right) -\psi _{5}\sin \left( wt\right) \right] \\ 
\\ 
-\left[ A\cos \left( Ay\right) -B\sin \left( Ay\right) \right] \left[ \psi
_{6}\cos \left( wt\right) -\psi _{7}\sin \left( wt\right) \right]%
\end{array}%
\right] ,  \TCItag{47}
\end{eqnarray}%
where we have used notations%
\begin{equation*}
\zeta _{i}=\frac{q_{i}(1+\alpha q_{i}+\gamma q_{i}^{2})^{2}}{%
(q_{i}^{2}+w^{2})(q_{i}-q_{4})(q_{i}-q_{5})},\text{ \ \ \ \ }\zeta
_{i}^{\ast }=\frac{w(1+\alpha q_{i}+\gamma q_{i}^{2})^{2}}{%
(q_{i}^{2}+w^{2})(q_{i}-q_{4})(q_{i}-q_{5})},i=1,2,3,
\end{equation*}%
\begin{eqnarray*}
\psi _{4} &=&\frac{\beta ^{2}(1+w^{2}\alpha -w^{2}\beta -w^{2}\gamma
+w^{4}\beta \gamma )}{1+w^{2}\left( 1+\beta \left( -2+w^{2}\beta \right)
\right) },\text{ \ \ \ }\psi _{5}=\frac{w\beta ^{2}(1-\alpha +w^{2}\alpha
\beta -w^{2}\gamma )}{1+w^{2}\left( 1+\beta \left( -2+w^{2}\beta \right)
\right) }, \\
\psi _{6} &=&\frac{\beta ^{2}}{\left[ 1+w^{2}\left( 1+\beta \left(
-2+w^{2}\beta \right) \right) \right] \left[ 1+w^{2}\left( \alpha
^{2}+\gamma \left( -2+w^{2}\gamma \right) \right) \right] } \\
&&\times \left[ \left( 
\begin{array}{c}
w^{8}\left( -2\alpha \beta ^{2}\gamma +2\beta \gamma ^{2}+a_{m}\beta \gamma
^{2}\right) +b_{m} \\ 
\\ 
+w^{2}\left( -2-a_{m}-c_{m}+2\alpha +2c_{m}\alpha -b_{m}\left( \left(
-2+\alpha \right) \alpha +\beta +2\gamma \right) \right) \\ 
\\ 
+w^{6}\left( 
\begin{array}{c}
-2\alpha ^{2}\beta -a_{m}\alpha ^{2}\beta +2\left( 1+a_{m}\right) \alpha
\gamma -4\beta \gamma -2a_{m}\beta \gamma -2\gamma ^{2} \\ 
-a_{m}\gamma ^{2}-c_{m}\gamma ^{2}-b_{m}\beta \gamma ^{2}+2\alpha \beta
\left( \beta +\left( 2+c_{m}\right) \gamma \right)%
\end{array}%
\right) \\ 
\\ 
w^{4}(-2\alpha -2a_{m}\alpha +2\alpha ^{2}+a_{m}\alpha ^{2}+c_{m}\alpha
^{2}+2\beta +a_{m}\beta -2(2+c_{m})\alpha \beta \\ 
\\ 
+2(2+a_{m}+c_{m})\gamma -2\alpha \gamma -2c_{m}\alpha \gamma +b_{m}(\alpha
^{2}\beta -2\alpha \gamma +\gamma \left( 2\beta +\gamma \right) ))%
\end{array}%
\right) \right. \\
&& \\
&&\left. +\left( \frac{%
\begin{array}{c}
\left( w^{2}\left( 1+a_{m}-\alpha \left( 1+c_{m}-w^{2}\beta \right) \right)
-\left( 1+a_{m}\right) w^{4}\gamma +b_{m}\left( -1+w^{2}\gamma \right)
\right) \\ 
\times \left( 1+w^{2}\left( \alpha ^{2}+\gamma \left( -2+w^{2}\gamma \right)
\right) \right) \left( 1+w^{2}\left( \alpha -\beta -\gamma \right)
+w^{4}\beta \gamma \right)%
\end{array}%
}{\left( 1+\alpha \left( -1+w^{2}\beta \right) -w^{2}\gamma \right) }\right) %
\right] , \\
&& \\
\psi _{7} &=&\frac{w\beta ^{2}}{\left[ 1+w^{2}\left( 1+\beta \left(
-2+w^{2}\beta \right) \right) \right] \left[ 1+w^{2}\left( \alpha
^{2}+\gamma \left( -2+w^{2}\gamma \right) \right) \right] } \\
&&\times \left[ 
\begin{array}{c}
\left( 1+w^{2}\left( \alpha -\beta \right) \right) \left( 1-2b_{m}\alpha
+w^{2}\left( 2+2a_{m}-\alpha \right) \alpha +\left( -1+w^{2}\alpha
^{2}\right) \left( -c_{m}+w^{2}\beta \right) \right) \\ 
\\ 
-w^{2}\gamma \left( 
\begin{array}{c}
2+2c_{m}+\left( 1+a_{m}\right) w^{2}-b_{m}\left( 1+\alpha \left(
2+w^{2}\left( \alpha -2\beta \right) \right) \right) \\ 
+w^{2}\left( \left( 1+a_{m}\right) \alpha \left( 2+w^{2}\alpha \right)
-2\left( 2+c_{m}+\left( 1+a_{m}\right) w^{2}\alpha \right) \beta
+2w^{2}\beta ^{2}\right)%
\end{array}%
\right) \\ 
\\ 
+w^{4}\gamma ^{2}\left( 
\begin{array}{c}
1-2b_{m}+w^{2}\left( 2+2a_{m}-\alpha -2\beta \right) +w^{4}\beta \left(
\alpha +\beta \right) \\ 
-c_{m}\left( -1+w^{2}\left( \alpha +\beta \right) \right)%
\end{array}%
\right) \\ 
\\ 
-w^{6}\left( -b_{m}+\left( 1+a_{m}\right) w^{2}\right) \gamma ^{3}%
\end{array}%
\right] . \\
&&
\end{eqnarray*}

\section{Graphical Results and Discussion}

To study the significant physical effects of the obtained results, the
impact of the material parameters on the fluid motion is highlighted by
graphical illustration of the velocity profiles for the flow due to the
sinusoidal oscillations of the infinite plate. To illustrate the difference,
we depicted the velocity profile for both cosine and sine oscillations of
the boundary. The numerical results are plotted for different values of time
t, magnetic parameter $M$, permeability parameter $K$, and the rheological
parameters $\beta $ and $\gamma $ of Burgers and generalized Burgers fluids.

Figure $1$ compares the profiles of velocity for different values of time
for cosine and sine oscillations of the boundary, respectively. It is noted
that the two oscillations have similar amplitudes and a phase shift that
persevere for all times. Here we can observe that as the bottom plate is set
into motion the velocity near the bottom plate is developing and fluctuating
around zero with the same frequency as the plate. Additionally, the fluid
oscillation has maximum amplitude adjacent the bottom plate and reduces far
away from the plate and approaches to zero.

Figure $2$ presents the profiles of the velocity for different values of the
magnetic parameter $M$ for both cosine and sine oscillations of the boundary
respectively. Since, magnetic field is applied in the transverse direction
and it is a force which resists the flow. Therefore, with the increase in
the values of the magnetic parameter $M$, the amplitude of the oscillation
tends to decrease. Also, a comparison shows that the amplitude of
oscillation is larger for hydrodynamic case ($M=0$) when compare with
hydromagnetic case ($M\neq 0$) case. The effects of the permeability
parameter $K$ are depicted in figure $3$. It has quite opposite effect on
the velocity profile to that of the magnetic parameter $M$. It is clearly
seen that with the increase of $K$ the velocity profile increases.

Figure $4$ shows the effect of the material parameter $\beta $ on the
velocity profile for both cosine and sine oscillations of the boundary,
respectively. We can clearly see from the figure that the velocity decreases
slightly by increasing the parameter $\beta .$ Further, the effect of the
rheological parameter $\gamma $ of the generalized Burgers fluid is shown in
figure $5$ for both cosine and sine oscillations of the boundary,
respectively. It is noted that an increase in the rheological \ parameter $%
\gamma $ of generalized Burgers fluid yield an effect opposite to that of
the parameter $\beta .$ From these figures, it is noticed that the profile
of the velocity for the sine oscillation are more sensitive compared to the
cosine oscillations of the boundary.

\bigskip

\begin{center}
\begin{tabular}{l}
\FRAME{itbpF}{272.7587pt}{191.5191pt}{0pt}{}{}{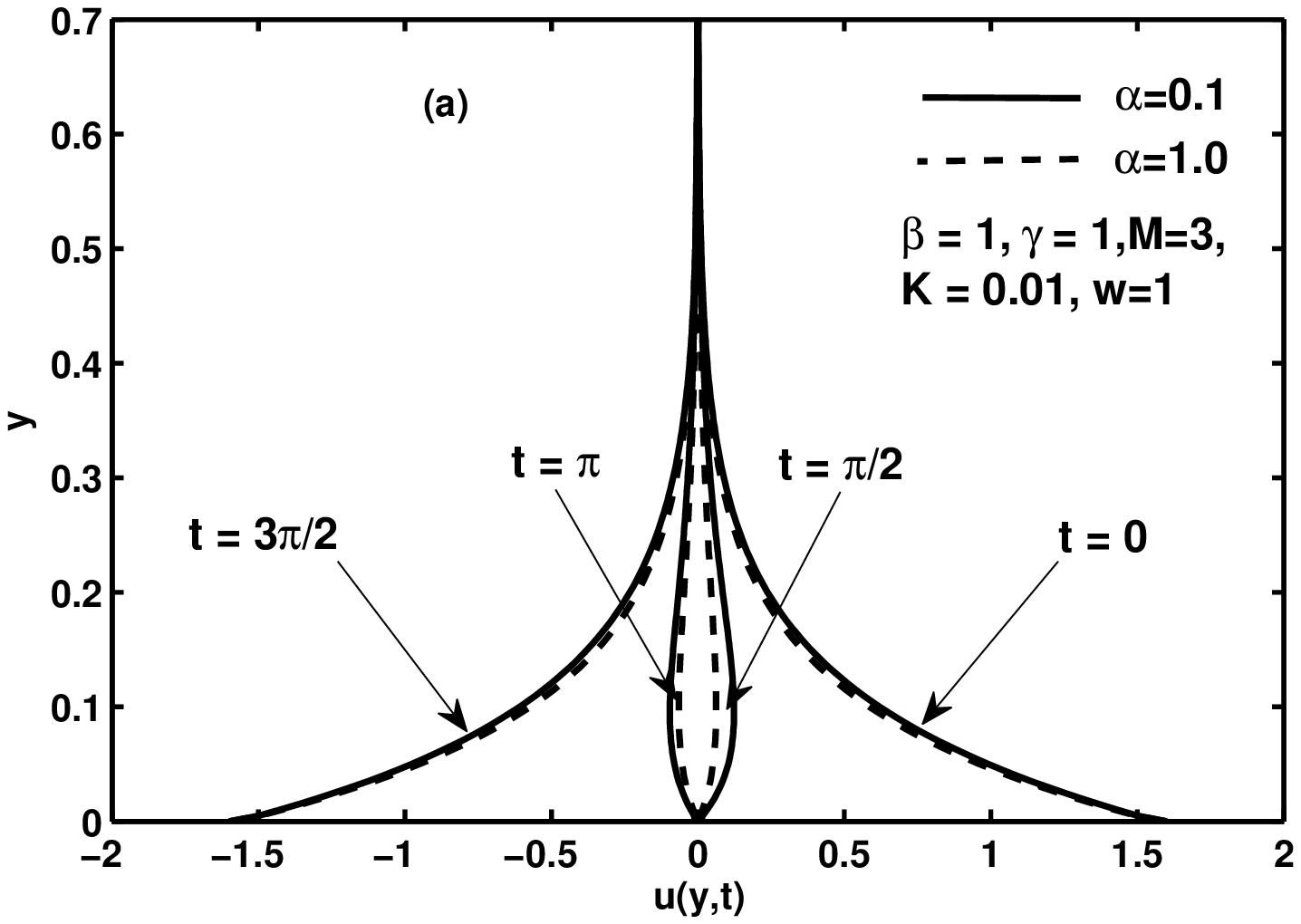}{\special{language
"Scientific Word";type "GRAPHIC";display "USEDEF";valid_file "F";width
272.7587pt;height 191.5191pt;depth 0pt;original-width
6.3976in;original-height 4.1652in;cropleft "0";croptop "1";cropright
"1";cropbottom "0";filename 'fig1a.eps';file-properties "XNPEU";}}\FRAME{%
itbpF}{272.7587pt}{191.5191pt}{0pt}{}{}{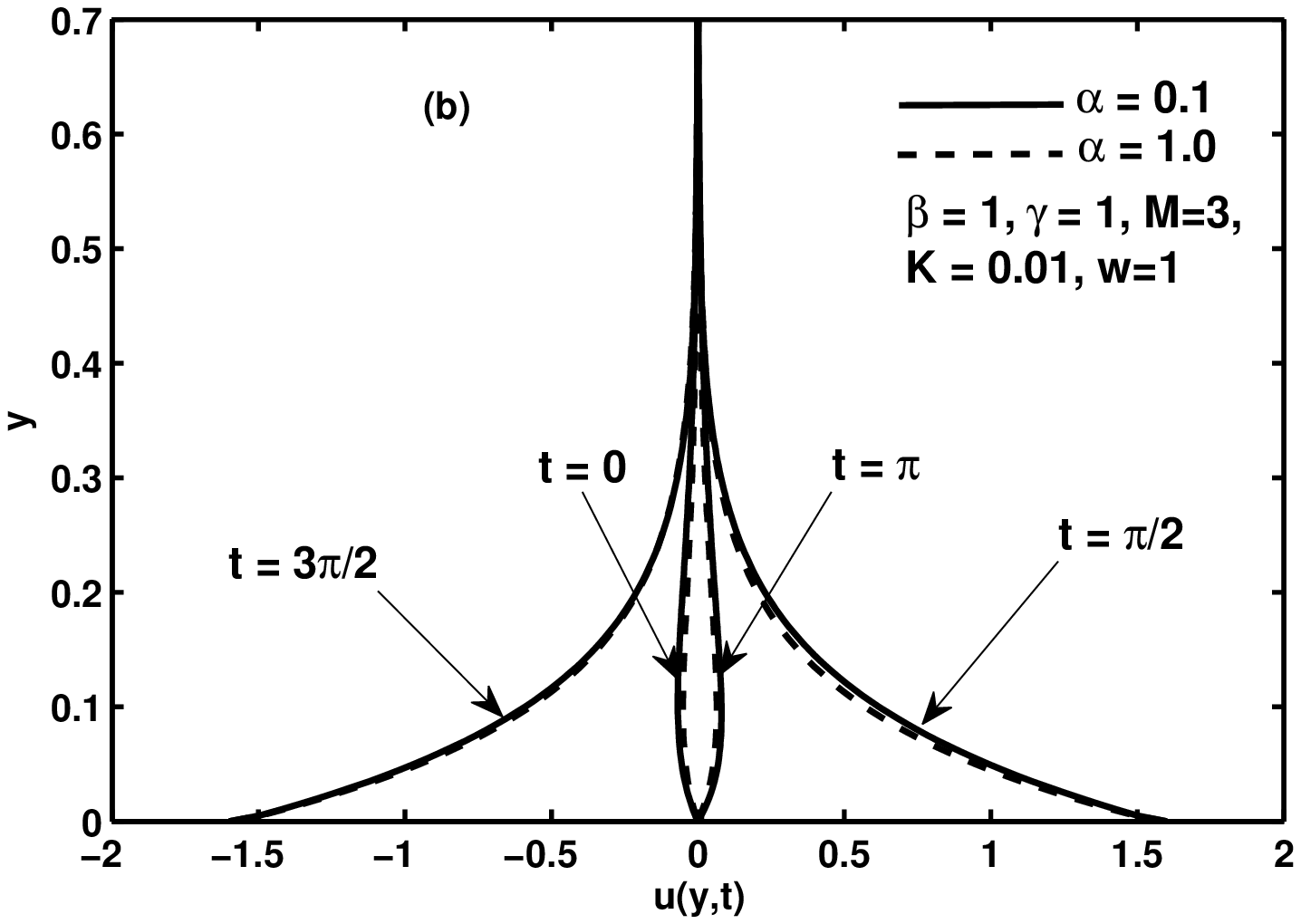}{\special{language
"Scientific Word";type "GRAPHIC";display "USEDEF";valid_file "F";width
272.7587pt;height 191.5191pt;depth 0pt;original-width
6.3976in;original-height 4.1652in;cropleft "0";croptop "1";cropright
"1";cropbottom "0";filename 'fig1b.eps';file-properties "XNPEU";}}%
\end{tabular}
\end{center}

Figure $1:$ Profiles of velocity $u\left( y,t\right) $ given by Eqs. $\left(
33\right) $ and $(34)$ for different values of time $t$ for cosine and sine
oscillations of the boundary, respectively$.$

\begin{equation*}
\FRAME{itbpF}{272.7587pt}{191.5191pt}{0pt}{}{}{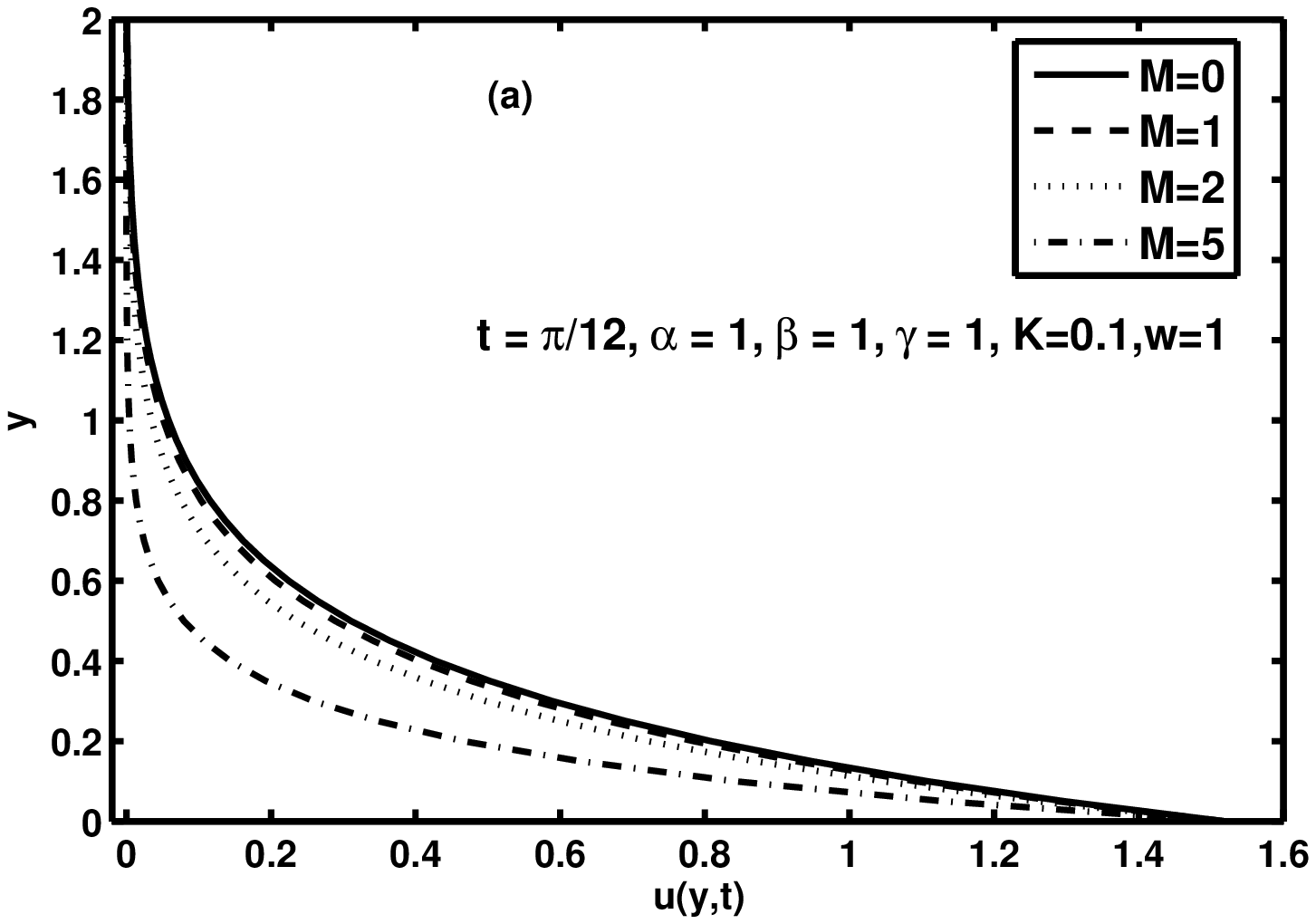}{\special{language
"Scientific Word";type "GRAPHIC";display "USEDEF";valid_file "F";width
272.7587pt;height 191.5191pt;depth 0pt;original-width
6.3976in;original-height 4.1652in;cropleft "0";croptop "1";cropright
"1";cropbottom "0";filename 'fig2a.eps';file-properties "XNPEU";}}\FRAME{%
itbpF}{272.7587pt}{191.5191pt}{0pt}{}{}{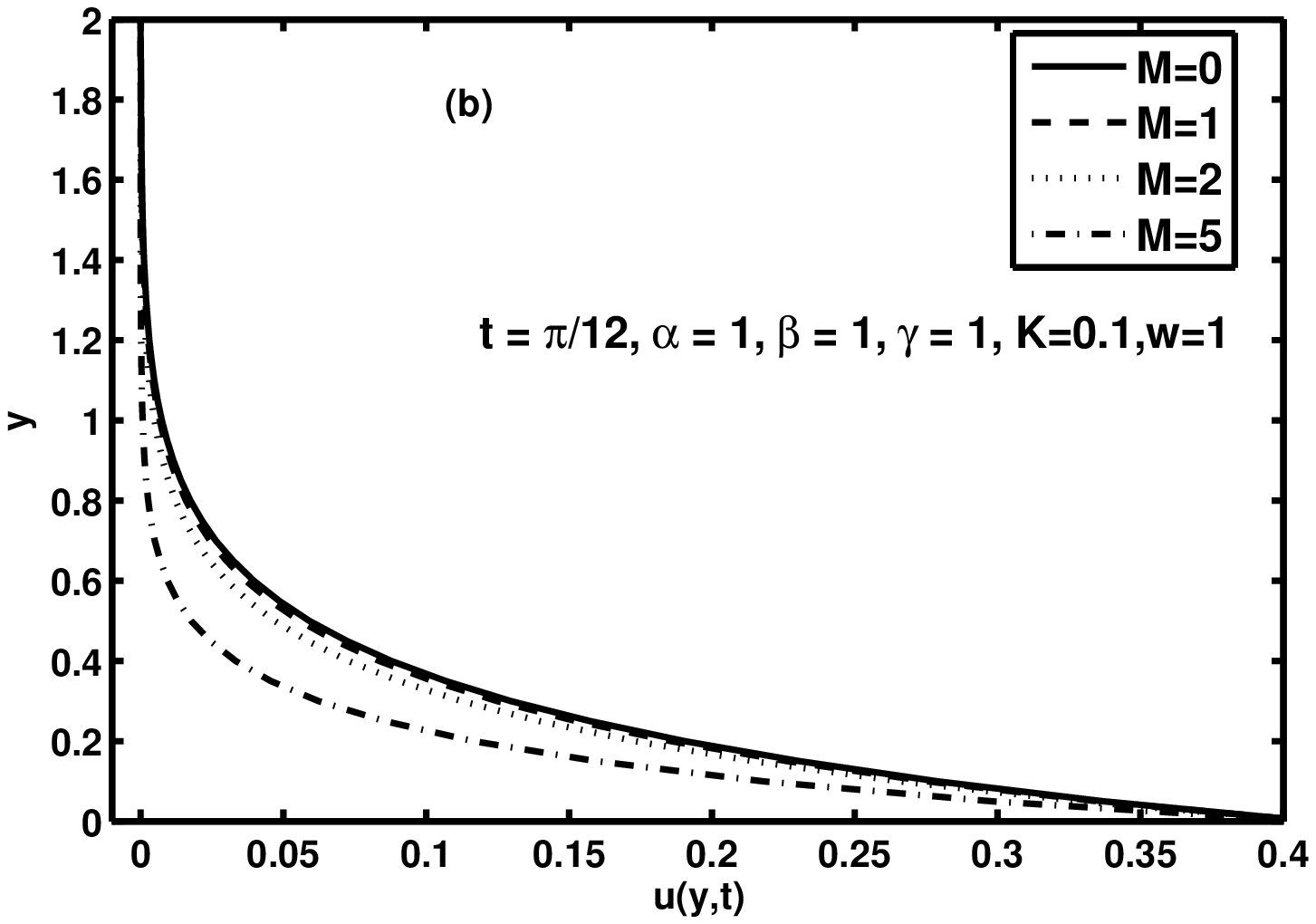}{\special{language
"Scientific Word";type "GRAPHIC";display "USEDEF";valid_file "F";width
272.7587pt;height 191.5191pt;depth 0pt;original-width
6.3976in;original-height 4.1652in;cropleft "0";croptop "1";cropright
"1";cropbottom "0";filename 'fig2b.eps';file-properties "XNPEU";}}
\end{equation*}

Figure $2:$ Profiles of velocity $u\left( y,t\right) $ given by Eqs. $\left(
33\right) $ and $(34)$ for different values of $M$ for cosine and sine
oscillations of the boundary, respectively$.$

\begin{equation*}
\FRAME{itbpF}{272.7587pt}{191.5191pt}{0pt}{}{}{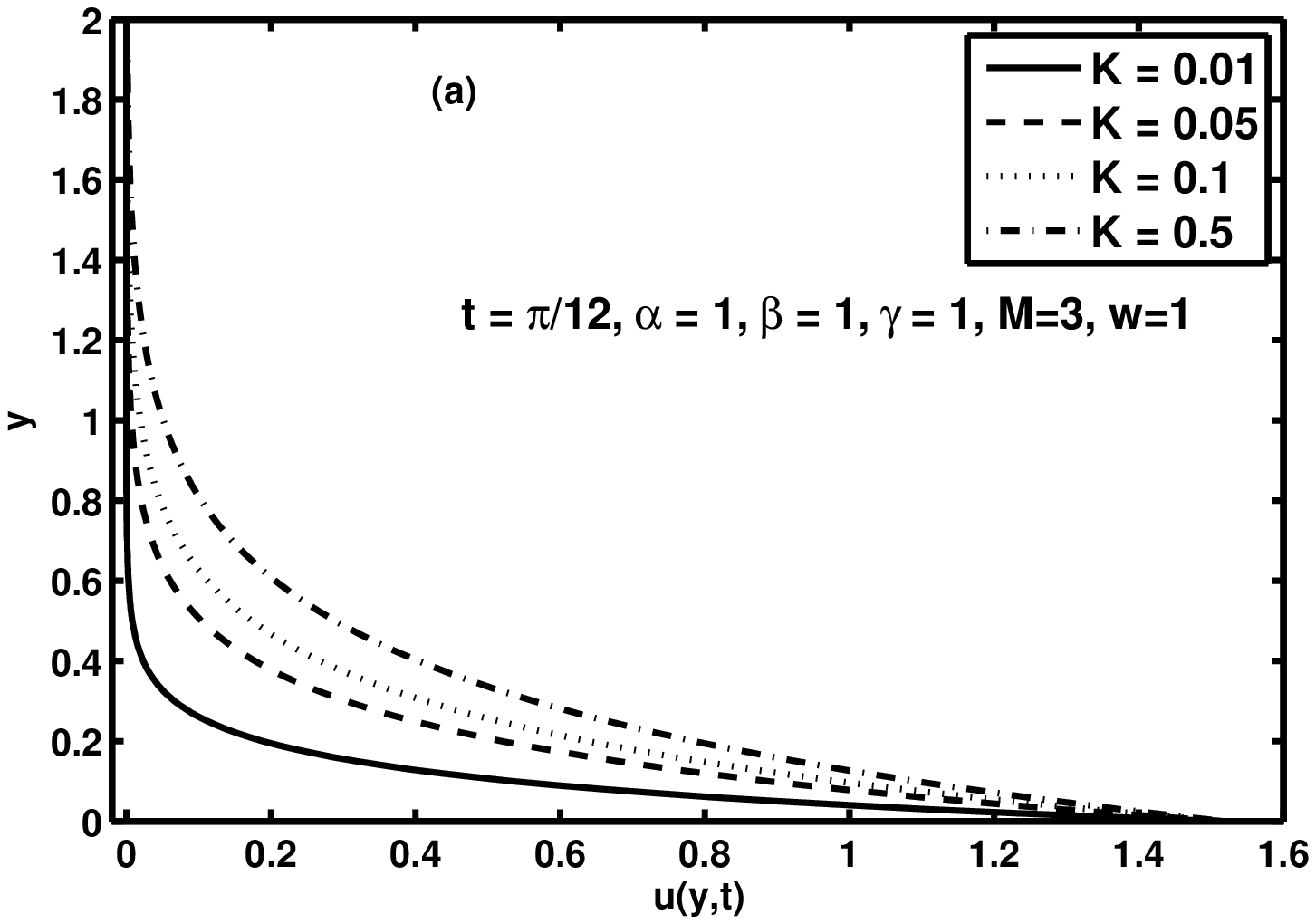}{\special{language
"Scientific Word";type "GRAPHIC";display "USEDEF";valid_file "F";width
272.7587pt;height 191.5191pt;depth 0pt;original-width
6.3976in;original-height 4.1652in;cropleft "0";croptop "1";cropright
"1";cropbottom "0";filename 'fig3a.eps';file-properties "XNPEU";}}\FRAME{%
itbpF}{272.7587pt}{191.5191pt}{0pt}{}{}{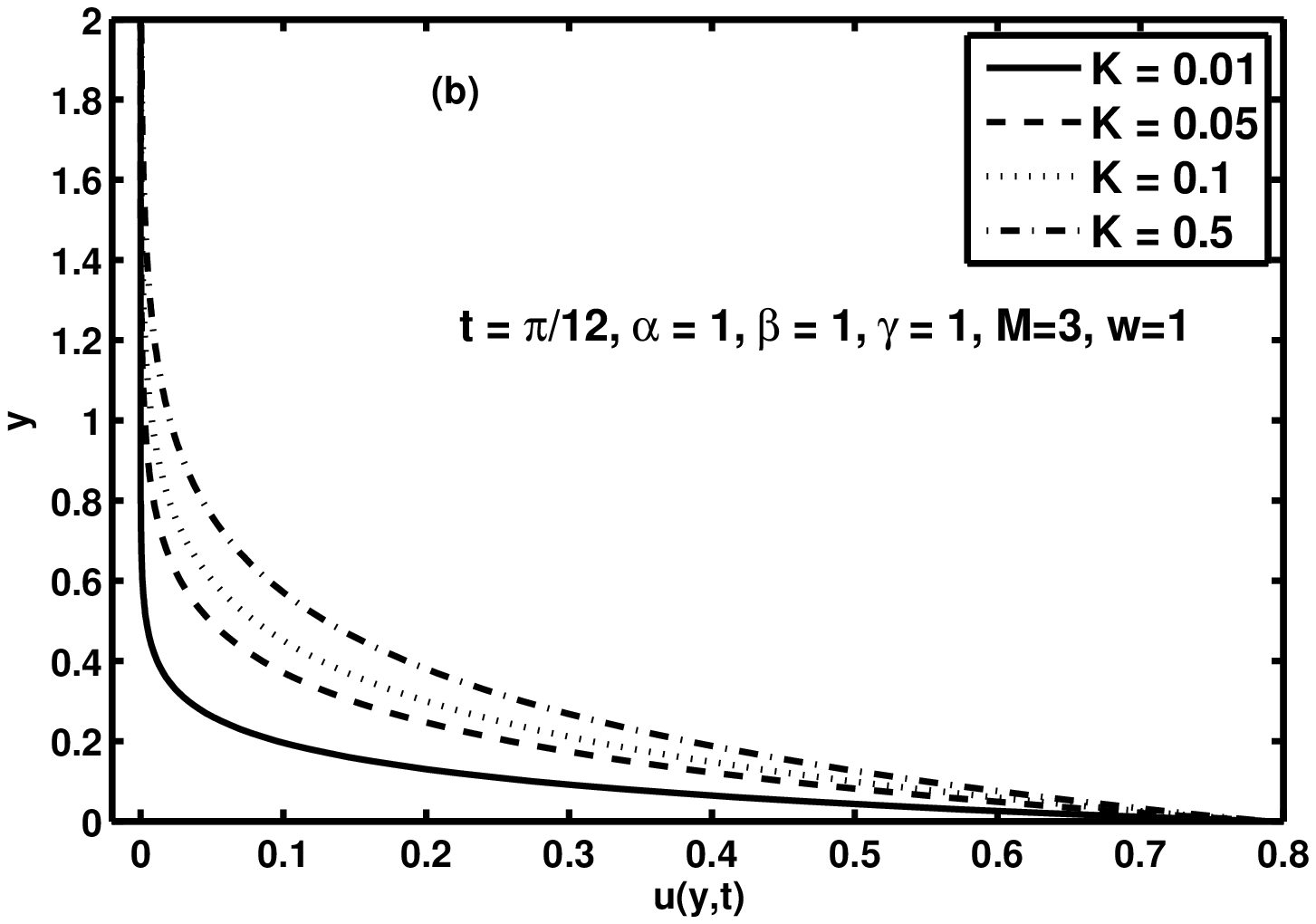}{\special{language
"Scientific Word";type "GRAPHIC";display "USEDEF";valid_file "F";width
272.7587pt;height 191.5191pt;depth 0pt;original-width
6.3976in;original-height 4.1652in;cropleft "0";croptop "1";cropright
"1";cropbottom "0";filename 'fig3b.eps';file-properties "XNPEU";}}
\end{equation*}

Figure $3:$ Profiles of velocity $u\left( y,t\right) $ given by Eqs. $\left(
33\right) $ and $(34)$ for different values of $K$ for cosine and sine
oscillations of the boundary, respectively$.$

\begin{center}
\bigskip\ \ \ 
\begin{tabular}{l}
\FRAME{itbpF}{272.7587pt}{191.5191pt}{0pt}{}{}{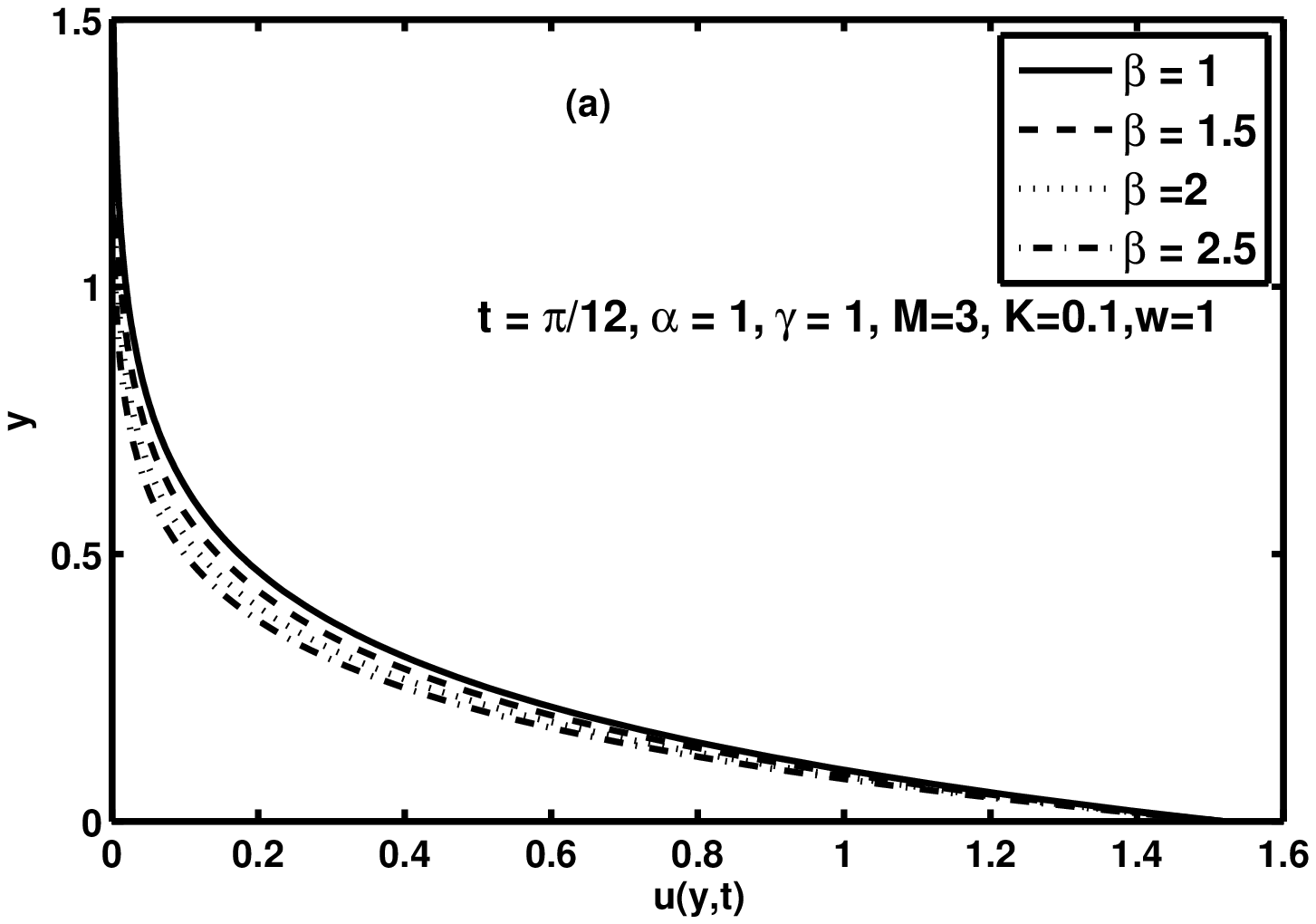}{\special{language
"Scientific Word";type "GRAPHIC";display "USEDEF";valid_file "F";width
272.7587pt;height 191.5191pt;depth 0pt;original-width
6.3976in;original-height 4.1652in;cropleft "0";croptop "1";cropright
"1";cropbottom "0";filename 'fig4a.eps';file-properties "XNPEU";}}\FRAME{%
itbpF}{272.7587pt}{191.5191pt}{0pt}{}{}{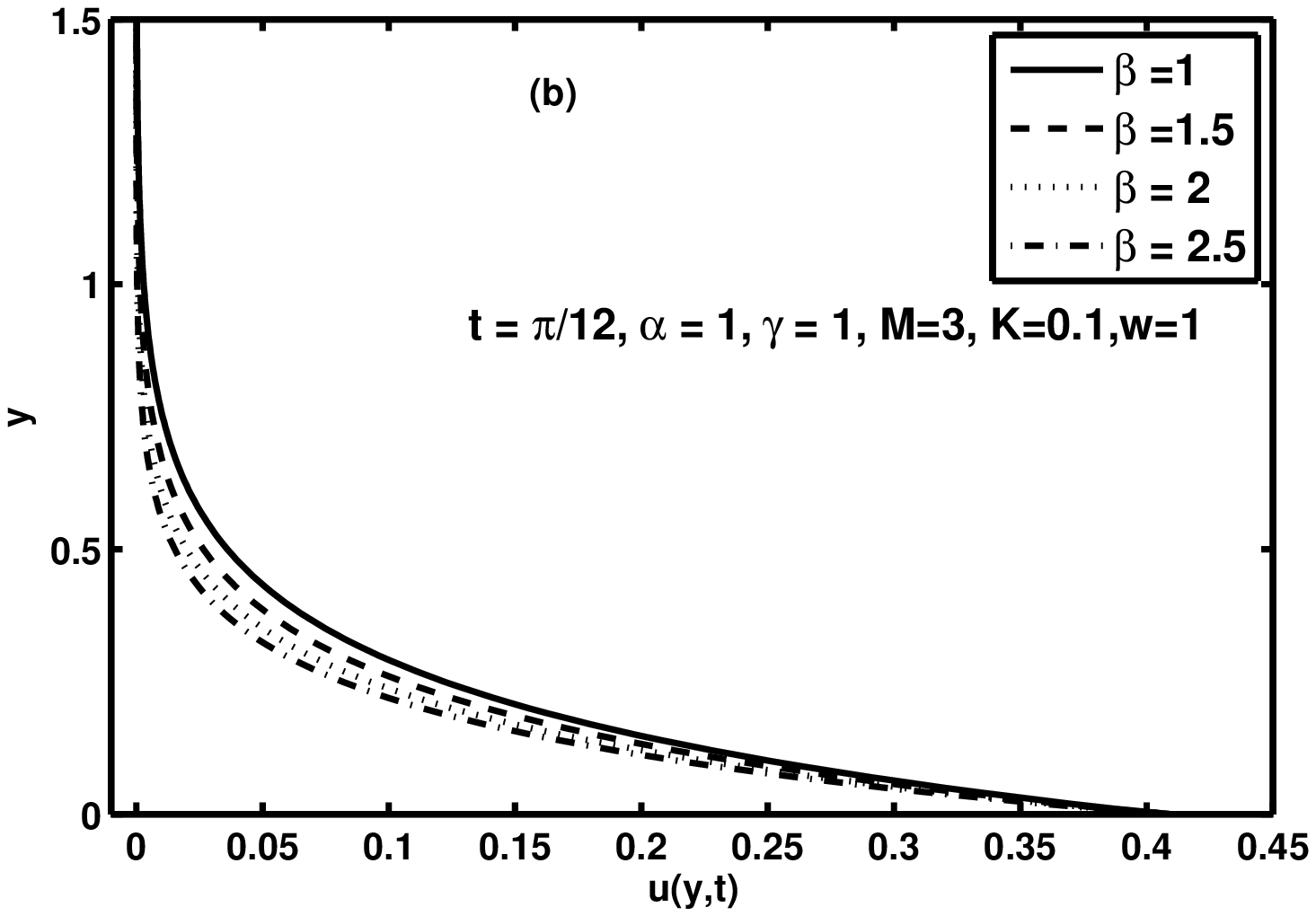}{\special{language
"Scientific Word";type "GRAPHIC";display "USEDEF";valid_file "F";width
272.7587pt;height 191.5191pt;depth 0pt;original-width
6.3976in;original-height 4.1652in;cropleft "0";croptop "1";cropright
"1";cropbottom "0";filename 'fig4b.eps';file-properties "XNPEU";}}%
\end{tabular}
\end{center}

Figure $4:$ Profiles of velocity $u\left( y,t\right) $ given by Eqs. $\left(
33\right) $ and $(34)$ for different values of $\beta $ for cosine and sine
oscillations of the boundary, respectively$.$

\begin{center}
\ 
\begin{tabular}{l}
\FRAME{itbpF}{272.7587pt}{191.5191pt}{0pt}{}{}{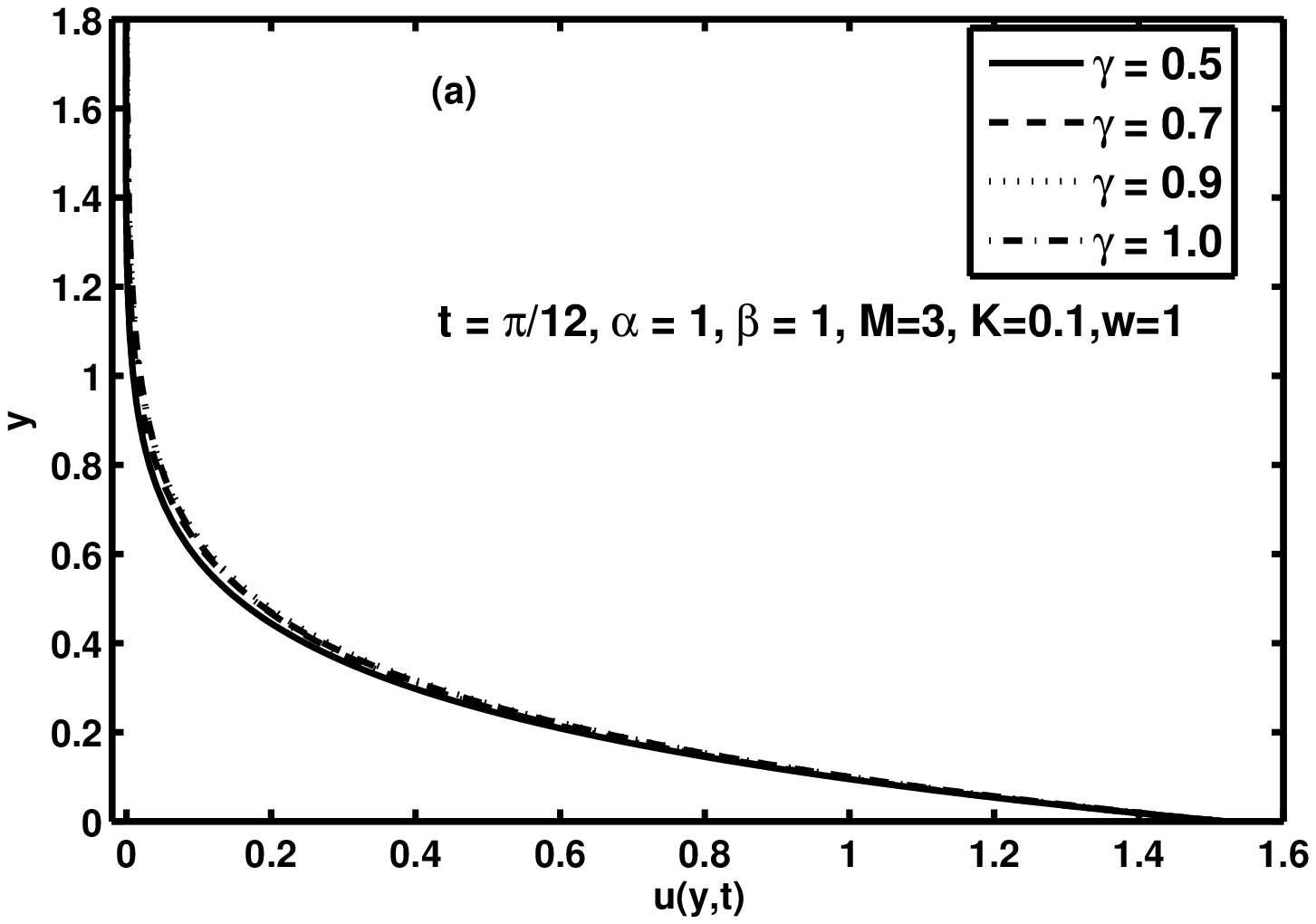}{\special{language
"Scientific Word";type "GRAPHIC";display "USEDEF";valid_file "F";width
272.7587pt;height 191.5191pt;depth 0pt;original-width
6.3976in;original-height 4.1652in;cropleft "0";croptop "1";cropright
"1";cropbottom "0";filename 'fig5a.eps';file-properties "XNPEU";}}\FRAME{%
itbpF}{272.7587pt}{191.5191pt}{0pt}{}{}{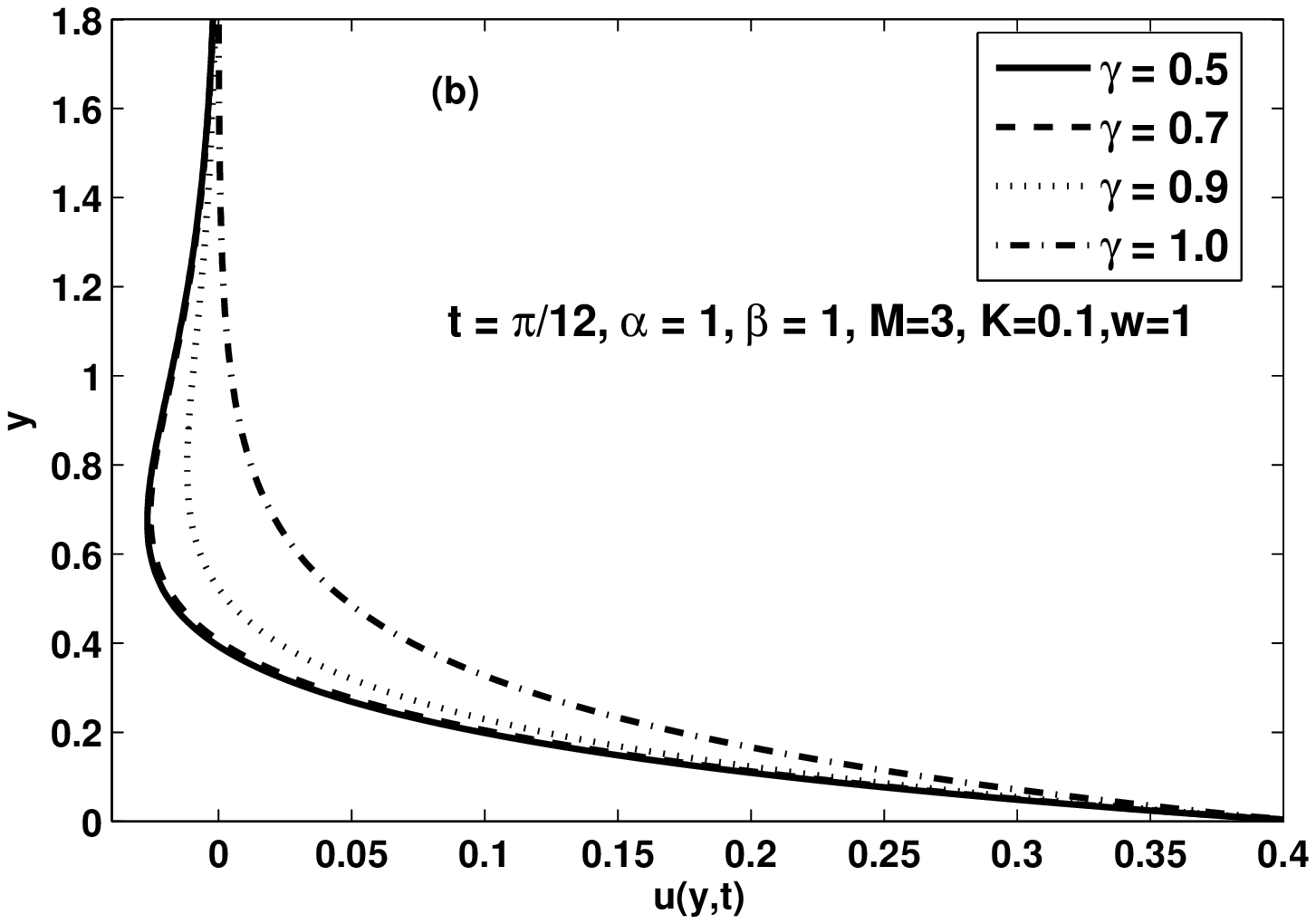}{\special{language
"Scientific Word";type "GRAPHIC";display "USEDEF";valid_file "F";width
272.7587pt;height 191.5191pt;depth 0pt;original-width
6.3976in;original-height 4.1652in;cropleft "0";croptop "1";cropright
"1";cropbottom "0";filename 'fig5b.eps';file-properties "XNPEU";}}%
\end{tabular}
\end{center}

Figure $5:$ Profiles of velocity $u\left( y,t\right) $ given by Eqs. $\left(
33\right) $ and $(34)$ for different values of $\gamma $ for cosine and sine
oscillations of the boundary, respectively$.$

\section{Brief Summary}

In this article the problem of an MHD flow through porous medium involving
generalized Burgers fluid has been discussed for cosine and sine
oscillations of the boundary. Modified Darcy's law for a generalized Burgers
fluid is used in the modelling of the governing equations. The results for
the flow are constructed by means of the Fourier sine and Laplace
transforms. These solutions, presented as a sum of the steady-state and
transient parts, explain the motion of the fluid for some times after its
initiation. After that time, when the transients disappear, the motion of
the fluid is described by the steady-state, which is independent of initial
conditions. Based on the solutions derived, we have analyzed the influence
of the various parameters on the velocity profile of the fluid.

\bigskip

\begin{quote}
\textbf{Acknowledgements: }This work has financial support of the Higher
Education Communication (HEC) of Pakistan.
\end{quote}


\bigskip

\begin{thebibliography}{99}
\bibitem{1} K. R. Rajagopal, Mechanics of non-Newtonian fluids, in: Recent
Development in Theoretical Fluids Mechanics, in: Pitman Research Notes in
Mathematics. Longman, New York, $291$ $(1993)$ $129-162.$

\bibitem{2} C. Fetecau, T. Hayat, M. Khan and Corina Fetecau, Unsteady flow
of an Oldroyd-B fluid induced by the impulsive motion of a plate between two
side walls perpendicular to the plate, Acta Mech., $198$ $(2008)$ $21-33.$

\bibitem{3} C. Fetecau, Corina Fetecau and D. Vieru, On some helical flows
of Oldroyd-B fluids, Acta Mech., $189$ $(2007)$ $53-63.$

\bibitem{4} W.C. Tan, W. Pan and M.Y. Xu, A note on unsteady flows of a
viscoelastic fluid with the fractional Maxwell model between two parallel
plates, Int. J. Non-Linear Mech., $38$ $(2003)$ $645-650.$

\bibitem{5} K.R. Rajagopal and R.K. Bhatnagar, Exact solutions for some
simple flows of an Oldroyd-B fluid, Acta Mech., $113$ $(1995)$ $233-239.$

\bibitem{6} N. Aksel, C. Fetecau and M. Scholle, Starting solutions for some
unsteady unidirectional flows of Oldroyd-B fluids, ZAMP, $57$ $(2006)$ $%
815-831.$

\bibitem{7} C. Fetecau and Corina Fetecau, Unsteady flows of Oldroyd-B
fluids in a channel of rectangular cross-section, Int. J. Non-Linear Mech., $%
40$ $(2005)$ $1214-1219.$

\bibitem{8} D. Vieru, C. Fetecau and Corina Fetecau, Flow of a generalized
Oldroyd-B fluid due to a constantly accelerating plate, Appl. Math. Comput., 
$201$ $(2008)$ $834-842.$

\bibitem{9} W.C. Tan and T. Masuoka, Stokes first problem for an Oldroyd-B
fluid in a porous half space, Phys. Fluids, $17$ $(2005)$ $023101.$

\bibitem{10} M. Khan, S.B. Khan and T. Hayat, Exact solution for the magneto
hydrodynamic flows of an Oldroyd-B fluid through a porous medium, J. Porous
Med., $10$ $(2007)$ $391-399.$

\bibitem{11} J. M. Burgers, Mechanical Considerations-model
system-phenomenological Theories of relaxation and of viscosity, in: J. M.
Burgers (Ed.), First Report on Viscosity and Plasticity, Nordemann
Publishing Company, New York, $(1935).$

\bibitem{12} J. M. Krishnan and K. R. Rajagopal, A thermodynamic framework
for the constitutive modeling of asphalt concrete: theory and application,
J. Mater. Civil Eng., $16$ $(2004)$ $155-166.$

\bibitem{13} A. R. Lee and A. H. D. Markwick, The mechanical properties of
bituminous surfacing materials under constant stress, J. SOC. Chem. Ind., $56
$ $(1937)$ $146-156.$

\bibitem{14} B. H. Tan, I. Jackson and J. D. F. Gerald, High-temperature
viscoelasticity of fine-grained polycrystalline olivine, Phys. Chem. Miner., 
$28$ $(2001)$ $641-664.$

\bibitem{15} W. R. Peltier, P. Wu and D. A. Yuen, The viscosities of the
earth mental, in: F. D. Stacey, M. S. Paterson, A. qicholas (Eds.),
Anelasticity in the Earth, American Geophysical Union, Colorado, $1981.$

\bibitem{16} T. Hayat, S.B. Khan and M. Khan, Influence of Hall current on
the rotating flow of a Burgers' fluid through a porous space, J. Porous
Med., $11$ $(2008)$ $277-287.$

\bibitem{17} T. Hayat, M. Hussain and M. Khan, Effect of Hall current on
flows of a Burgers' fluid through a porous medium, Transp. Porous Med., $68$ 
$(2007)$ $249-263.$

\bibitem{18} M. Khan, S.H. Ali and C. Fetecau, Exact solutions of
accelerated flows for a Burgers' fluid, I. The case $\gamma <\lambda ^{2}/4,$
Appl. Math. Comput., $203$ $(2008)$ $881-894.$

\bibitem{19} T. Hayat, C. Fetecau and S. Asghar, Some simple flows of a
Burgers' fluid, Int. J. Eng. Sci., $40$ $(2006)$ $1423-1431.$

\bibitem{20} P. Ravindran, J.M. Krishnan and K.R. Rajagopal, A note on the
flow of a Burgers' fluid in an orthogonal rheometer, Int. J. Eng. Sci., $42$ 
$(2004)$ $1973-1985.$

\bibitem{21} T. Hayat and M. Sajid, Homotopy analysis of MHD boundary layer
flow of an upper-convected Maxwell fluid, Int. J. Eng. Sci., $45$ $(2007)$ $%
393-401.$

\bibitem{22} M. Khan, T. Hayat and S. Asghar, Exact solutions of MHD flow of
a generalized Oldroyd-B fluid with modified Darcy's law, Int. J. Eng. Sci., $%
44$ $(2006)$ $333-339.$

\bibitem{23} M. Khan, S.H. Ali, C. Fetecau and T. Hayat, MHD flows of a
second grade fluid between two side walls perpendicular to a plate through a
porous medium, Int. J. Non-Linear Mech., $43$ $(2008)$ $302-319.$

\bibitem{24} T. Hayat, S.B. Khan and M. Khan, Exact solution for rotating
flows of a generalized Burgers' fluid in a porous space, Appl. Math. Model., 
$32$ $(2008)$ $749-760.$

\bibitem{25} T. Hayat, M. Khan and S. Asghar, On the MHD flow of fractional
generalized Burgers' fluid with modified Darcy's law, Acta Mech. Sin., $23$ $%
(2007)$ $257-261.$

\bibitem{26} I.N. Snedden, Fourier Transforms, McGraw Hill Book Company,
Inc., New York-Toronto-London, $1951.$
\end{thebibliography}
\end{document}